# Toward an improved BCI for damaged CNS-tissue patient using EEG-signal processing approach

Fateme Dehrouye-Semnani[1], Nasrollah Moghada Charkari[2], Seyed Mohammad Mehdi Mirbagheri[3*]

[1]Medical Informatic Department, Faculty of Medical Sciences, Tarbiat Modares University, Tehran, Iran, fateme.dehrouye@modares.ac.ir

[2]Department of Computer Engineering, Faculty of Electrical and Computer Engineering, Tarbiat Modares University, Tehran, Iran, charkari@modares.ac.ir

[3*]Seyed Mohammad Mehdi Mirbagheri, Northwestern University/Rehabilitation Institute of Chicago, USA, mehdi@northwestern.edu

## Abstract

This article examined brain signals of people with disabilities using various signal processing methods to achieve the desired accuracy for utilizing brain-computer interfaces (BCI). EEG signals resulted from 5 mental tasks of word association (WORD), Mental subtraction (SUB), spatial navigation (NAV), right-hand motor imagery (HAND), and feet motor imagery (FEET) of 9 people with central nervous system (CNS) tissue damage were used as input. In processing this data, Butterworth band-pass filter (8-30 Hz) was used in the preprocessing, and CSP, TRCSP, FBCSP methods were used in feature extraction, and LDA, KNN, linear and nonlinear SVM were used in classification stages. The training and testing process was repeated up to 100 times, and the random subsampling method was used to select the training and test data. Mean accuracy in 100 replications was reported as final accuracy. The classification results of two classes of 5 mental tasks on the data of 9 people with central nerve damage showed that the combination of FBCSP with KNN classifier has the highest accuracy of $69\pm1.1$ which belongs to the two classes of word association and feet motor imagery. The study of different feature extraction methods and classification indicates that the proper feature selection method and especially classifier is crucial in accuracy rate. Furthermore, it is necessary to pay attention to successful signal processing methods in designing a brain-computer interface to use in people with central nervous system.

# 1. Introduction

Over the years, one of the most important issues in the world has been the problems of people with mobility and speech disabilities who have suffered from central nervous system (CNS) tissue damage as a result of accidents and diseases [1]. Brain-Computer Interfaces (BCIs) seek to improve these individuals' living conditions by converting mental tasks into control commands through a non-muscular pathway [2]. For BCI to be feasible, brain activities must be classified appropriately. As a result, the primary and essential element in the implementation of such systems is the accuracy of the classification of brain activity or Electroencephalogram (EEG) related to individuals [3].

Central nervous system diseases are a group of neurological disorders that affect the brain or spinal cord's structure or function due to factors such as trauma and etc. [4-6]. Spinal cord injury and stroke are complications of central nerve damage that cause muscle loss and paralysis. The prevalence of spinal cord injury worldwide is between 40 and 80 per million. Approximately 250,000 to 500,000 people worldwide suffer from spinal cord injury each year [7]. Also, 15 million people worldwide suffer from stroke each year, of which 5 million have a disability [8]. Since the number of people with central nervous system tissue damage is high and their maintenance requires a lot of time and money, it is important to pay attention to the solutions developed to improve the living conditions of these people.

Since the last decade, serious efforts have been made to connect computers to the environment based on brain electrical signals analysis. Recent advances in neuroscience, physiology, signal processing and analysis, machine learning, and hardware technology have made it possible to design direct brain-computer interfaces, abbreviated as BCI, so that physically disabled patients can meet their personal needs [9]. BCI is a device that translates brain activities into instructions and sends them to a computer or other devices.

BCI makes it possible for people with neurological disorders, spinal cord injuries, and severe motor disabilities to interact with the outside world through a non-muscular pathway [2]. The BCI system operates in three axes of signal acquisition, signal processing, and system output [10]. Activities selected for classification should be easy for training the user, and the system can classify them correctly [11]. First, brain signals must be recorded in the system using brain signal recording devices [12]. Brain signal recording is divided into three categories: Invasive, semi-invasive, and Non-invasive. In the invasive category, signals from neurosurgery are used; in the semi-invasive category, Electrocardiography (ECOG) signals, and in the non-invasive category, EEG, Magnetoencephalography (MEG), and Functional magnitude radiography signals (fMRI) are used in BCI systems. Usually, due to high time accuracy and Low cost, and ease of use, EEG is used to record brain signals in BCI systems [2].

The processing of EEG signals plays a vital role in the performance of systems that follow these signals. Also, in the BCI field, the processing of EEG signals resulting from mental tasks determines such systems' effectiveness. EEG signal processing is done in 3 stages of preprocessing, extraction and selection of features, and classification [13]. Signal preprocessing methods are mainly aimed at removing noise from noisy signals and increasing the signal-to-noise ratio to extract relevant and important information from the signals. Therefore, signal preprocessing is required before feature extraction. The preprocessing of EEG signals for BCI applications is mainly done by frequency domain filters such as band-pass filter along with other filtering techniques [13]. After this step, useful signal properties representing the users' intellectual activity represent the preprocessed brain signals [13].

Over the years, Fourier transform methods and related methods, wavelet transform, principal component analysis, independent component analysis, autoregressive methods, or a combination of these methods have been used to extract features from EEG signals. Among them, Fourier transform is one of the most popular ways of feature extraction in this field [14]. In recent years, the advent of CSP and related methods has shown that it is a more powerful method than Fourier transform for feature extraction [15]. Yijun Wang et al. [16] used the CSP feature extraction method along with LDA classification. Kai Keng Ang et al. [17] used the FBCSP method along with the SVM classifier.

Fabien Lotte et al. [18] used CSP regulated methods as RCSP with LDA classifier. Among the RCSP algorithms for EEG signals classifying, the CSP algorithm with Tokhof Regulated and Tokhof weight Regulated has been one of the best feature extraction algorithms. Classification of EEG signals uses classifiers such as LDA, SVM (linear and nonlinear), neural networks, nonlinear Bayesian classifiers, nearest-neighbor classifiers, and a combination of classifiers. LDA and SVM classifiers are among the most popular classifiers in this field [3]. Al Zoubi et al. [19] used the power spectrum method for feature extraction and SVM, KNN, and Bayesian approaches to classify EEG signals; It was shown that the KNN classifier obtained higher accuracy than other classifiers. The classification of brain activities can also be used in virtual reality systems [20], the Internet of Things [21], various types of brain-computer interfaces such as communication, movement, control, rehabilitation, etc. [12]. Reinhold Scherer et al. [22] used the CSP feature extraction method along with the LDA classifier to analysis the mental tasks of people with central nervous system (CNS) tissue damage to classify five mental tasks in the BCI field. Hence, this paper seeks to increase the accuracy of binary classification of mental tasks of people with central nervous system (CNS) tissue damage to use in BCI system using CSP, FBCSP, and TRCSP methods for feature extraction and KNN and LDA and linear and nonlinear SVM for classification.

## 2. Materials and methods

### 2-1 Data and pre-processing

The data set used in this article was obtained from the BNCI Horizon2020 database, which is a freely accessible database. Database No. 13 from the BNCI Horizon2020 database includes nine people with severe mobility impairments (six women and three men).

Participants A and C suffer from locked-in syndrome, numbers D, E, G, J, L suffer from spinal cord injury C5, C4 with grades A, C, and participants F, H suffer from hemorrhagic stroke. The details of the participants are summarized in table 1.

table1- Details of participants from the BNCI Horizon2020 [22]

| Participant details | | | | |
|---|---|---|---|---|
| ID | Gndr | Age | Mth | Event |
| A | M | 42 | 6 | Locked-in syndrome due to brainstem stroke |
| C | F | 31 | 5 | Locked-in syndrome due to brainstem stroke |
| D | F | 33 | 2 | Spinal cord injury C5, ASIA C |
| E | F | 40 | 255 | Spinal cord injury C5, ASIA A |
| F | F | 57 | 5 | Massive hemorrhagic stroke in left hemisphere |
| G | F | 43 | 27 | Spinal cord injury C5, ASIA C |
| H | F | 20 | 6 | Hemorrhagic stroke parietotemporal, right central no cranium |
| J | M | 36 | 53 | Spinal cord injury C5, ASIA A |
| L | M | 38 | 15 | Spinal cord injury C4, ASIA A |

EEG signals were recorded from 30 electrode channels placed on the scalp following International System 10-20. Besides, electro conductivity activity was recorded from two electrodes placed on the outer canal of the left eye—GAMMAsys g.tec system with active g. LADYbird electrodes and two bio-signal amp amplifiers, Graz (g.USB, Guger Technologies), were used to record signals. The signals were filtered through a 0.5-100 Hz band-pass filter (notch filter at 50 Hz) and sampled at 256 Hz. The experiments were performed on two different days with at least five days between two weeks. Details of the registration process are summarized in Figure 2-2. In each session, eight mental tasks were performed for each class, which resulted in 40 experiments from each class for each day. Each experiment composed of 25 signs, which were about five mental tasks. Symptoms were presented in random order. Participants were asked to perform a continuous mental imagery task for 7 seconds. Mental tasks include word association (WORD), mental subtraction (SUB), spatial navigation (NAV), right-hand motor imagery (HAND), and feet motor imagery (FEET). Participants were placed approximately 0.7 m in front of a 17-inch computer monitor. Participants received written and oral instructions for each task before the start of each trial. They were asked to perform the desired activities in relaxation and avoiding movements, and reducing blinking during the trial. Individuals were asked to perform an exercise consisting of two experiments in each class before recording the experiments to familiarize themselves with the experimental pattern. The duration of the trial was 10 seconds. Initially, at t = 0, a cross was displayed in the middle of the screen. Participants were asked to loosen and secure the cross to avoid eye movements. At t = 3 seconds, a beep sounded to grab the participant's attention. The symbol representing the requested illustration task is one of the five graphic symbols, from t = 3 sec. to t = 4.2 sec. At t = 10 sec, a second beep sounded, and the stabilizing cross disappeared, signaling the end of a trial. The rest (interval between re-tests) occurs between 2.5 seconds and 3.5 seconds before starting the next test. Participants were asked to refrain from movements of the head during the filming and to blink during rest. The registration steps are shown in Figure 1.

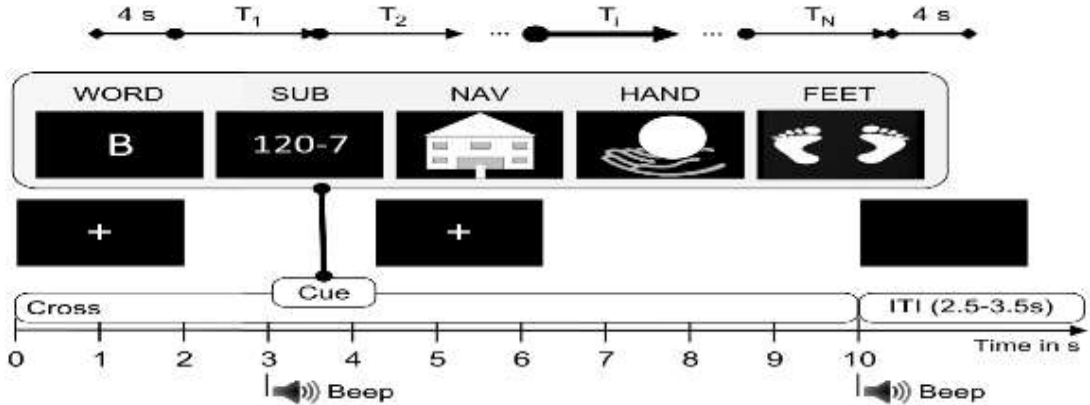

Figure 1: Experimental registration steps [22]

In general, EEG signals contain noise and artifacts as described in the previous Section. In practice, an 8-30Hz Band-pass filter using a Butterworth filter in its design is used to improve the signal quality and remove the EEG signal artifacts.

## 2-2. Feature Extraction

In the extraction step, the spatial filtering method is very effective for separating motor imagery EEG signals. In this section, we review the spatial filters available for this application. The CSP algorithm and other extended CSP algorithms such as TRCSP and FBCSP are discussed below.

### 2-2-1. Common Spatial Pattern Algorithm

The CSP algorithm was first proposed as a feature classification algorithm in [23]. It was initially used to diagnose disorders in clinical EEG [24]. Later in an experiment, CSP was used to distinguish two classes of motion. A similar study was performed to separate motor imagery from multichannel EEG data. Recorded EEG signals are considered as

$$X(t) = [x_1(t), x_2(t), \ldots, x_n(t)]^T \quad (1)$$

that use n channels during the imagining of right and left hands movements. The purpose of the CSP algorithm is to compute spatial filters that separate two motor imagery perceptions by maximizing the rate of variance of signals between two classes. The CSP objective function is defined as follows

$$\min_w \backslash \max_w J(w) = \frac{w^T Cov_1 w}{w^T Cov_2 w} \quad (2)$$

Where $w \in R^n$ is the space filter for optimization, Cov1 and Cov2 indicate the mean values of class1 and class2 covariance matrices. The covariance matrix x (t) with a mean of zero for class C is given in 3;

$$Cov_c = \frac{1}{T}\sum_{t=1}^{T}(x(t)-\bar{x})(x(t)-\bar{x})^T \quad c \in \{1,2\} \quad (3)$$

The result of Equation 2 by solving the vertical numerical division is in Equation 4;

$$Cov_1 w_1 = \lambda Cov_2 w_1 \quad (4)$$

Where $\lambda$ represents the eigenvalue. Special vectors are arranged according to their separability. The eigenvector with the largest eigenvalue appears to have the highest resolution compared to the rest. Spatial filters $W=[w_1.....w_p]$ (which is the set of all selected filters) for both classes are obtained by $P/2$ special vectors with the largest and smallest eigenvalues [25].

Let's the bandpass-filtered EEG sample $x_n \in R^{C \times T}$, where n is the nth sample, C is the number of channels and T is the number of time points. The spatially filtered signal $Z_n$ is found as follows

$$Z_n = W_{CSP}^t X_n \quad (5)$$

Where, $W_{CSP}$ is the CSP spatial filter making by selecting the first and the last m columns of CSP matrix, W. The variance-based CSP features of nth sample are then extracted using as follows

$$F_n^i = \log\left(\frac{var(Z_n^i)}{\sum_{j=1}^{2m} var(Z_n^j)}\right) \quad (6)$$

Where, $F_n^i$ denotes the ith feature of the nth sample, and $var(Z_n^i)$ is the variance of the $j^{th}$ row of $Z_n$ $Z_n$. Using the extracted features of each sample, the feature set is constructed as follows

$$F = \begin{bmatrix} F_1^1 & \cdots & F_1^{2m} \\ \vdots & \ddots & \vdots \\ F_N^1 & \cdots & F_N^{2m} \end{bmatrix} \quad (7)$$

Where, N is the number of samples. So, the features of training and test samples are used to make the training and test feature sets [26].

### 2-2-2. Tokhof regularized standard spatial algorithm

The CSP algorithm calculates spatial filters using a covariance matrix. The presence of outdated data leads to poor classification performance. One way to reduce the likelihood of such events is to add prior information to the CSP optimization problem in the form of regulatory statements. In general, prior information can be used in two steps. The data can be executed in the covariance matrix estimation step or they can be entered directly into the objective function of the CSP algorithm. To do this, various regularization methods have been proposed. Tuning is one of the most common methods used in machine learning to develop a change-resistant system. The reset of the CSP is mainly done by estimating the covariance matrix or by using the penalty value of the objective function [18]. The configured CSP objective functions can be represented as

$$\tilde{J}_1(w) = \frac{w^T \widetilde{Cov}_1 w}{w^T \widetilde{Cov}_2 w + \eta P(w)} \tag{8}$$

$$\tilde{J}_2(w) = \frac{w^T \widetilde{Cov}_2 w}{w^T \widetilde{Cov}_1 w + \eta P(w)} \tag{9}$$

Where p is the penalty term, r is the setting parameter, and Cov is the estimated covariance matrix of class c.

The RCSP model itself has different types [18]. One of these regularization methods is the TRCSP method with Tokhof regularization in which the regularization is performed on the objective function. The objective function is defined as follows

$$\tilde{J}_{p1}(w) = \frac{w^T C_1 w}{w^T C_2 w + \alpha P(w)} \tag{10}$$

p (w) is penalty function indicates in which extent filter w contains previous information. P (w) must be minimized to maximize $\tilde{J}_{p1}(w)$. The parameter α is positive and determines the importance of the penalty function and is determined by the user. In [18] the penalty function is defined as $P(w) = w^T K w$. This function allows the obtained filters to have a small softness and reduces the effect of interference and unusual samples. The space filters that maximize the above relation are the special vectors corresponding to the largest eigenvalues of the matrix $M_1 = (\tilde{C}_2 + \alpha K)^{-1} \tilde{C}_1$ and $M_2 = (\tilde{C}_1 + \alpha K)^{-1} \tilde{C}_2$. In the Tokhof method, the matrix k in the penalty function is defined as k = I. It has been shown that the Tokhof regularized common spatial algorithm performs better than other regularized spatial pattern methods [18].

### 2-2-3. Filter Bank Common Spatial Pattern

In the shared spatial filter method, instead of the frequency filters being optimized simultaneously with the spatial filters to find the optimal band of each subject, the signal is first filtered by a spatial filter under different frequency ranges. Then in each interval, spatial filters are learned by the CSP algorithm. Finally, out of all the obtained features, the best features are selected through feature selection algorithms. Accordingly, both spatial and frequency filters are selected simultaneously since each attribute fits a frequency sub-band and a spatial filter. Figure2 shows the steps of this method. The FBCSP method has shown outstanding results since its introduction and has had higher accuracy and better performance on the BCI competition database than other algorithms [12].

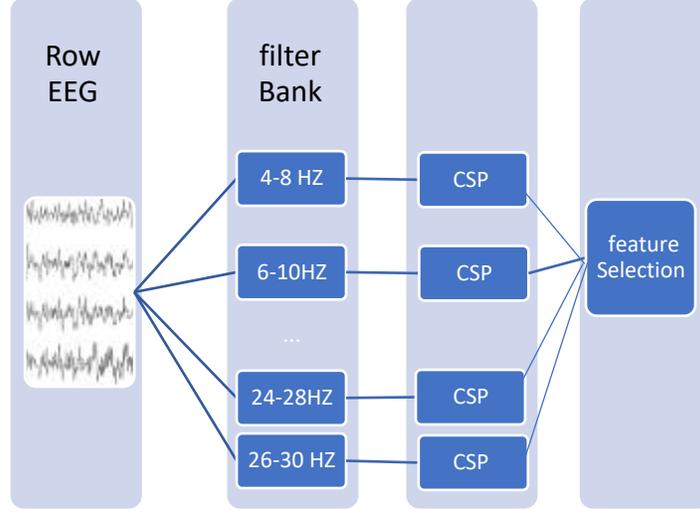

Figure 2: Steps of the FBCSP Method

## 2-3. Classification

After the feature extraction step, the classification should be done, in which the extracted features are classified to determine which class they belong to using classification algorithms. The different types of BCI classifications are linear classification [27], nonlinear classification [28], neural networks [29], nearest neighbor [30], and ensemble of classifiers [31]. Among these classification methods, linear classification is usually used to separate motor imagery where popular linear classifiers for motor imagery signals include linear dimensional analysis and support vector machine [20]. It has been found that the nearest neighbor classifier performs better compared with other classifiers where data is not high dimension [3].

### 2-3-1. Linear dimensional analysis

The main purpose of the LDA is to classify multidimensional data into a reduced dimensional subspace with higher resolution. The LDA method mainly considers the data of each class as a model of the probability density function. The class of each input data is separated from other classes using the largest possible value of the density function. The LDA assumes that all classes have a normal distribution and a similar covariance matrix. There is a class like C and $X = [x_1, \cdots, x_n]^T$ are instances for classification where n represents the number of instances. The mean, $\bar{x}$, and the covariance matrix, Cov can be represented as follows

$$\bar{x}_c = \frac{1}{n_k} \sum_{i=1}^{c} x_i \tag{11}$$

$$Cov_c = \frac{1}{n_c} \sum_{i=1}^{c} (x_t - \bar{x}_i)(x_t - \bar{x}_i)^T \tag{12}$$

The x data points are then categorized as;

$$g(x) = \arg\max_c x_t Cov_c^{-1}\bar{x}_c - \frac{1}{2}\bar{x}_c^T Cov_c^{-1}\bar{x}_c \qquad (13)$$

Where the decision boundary is a linear function. Class x is defined by the objective function in Equation 9. LDA is mainly used for binary classification, but can also be used for multi-class problems [25].

**2-3-2. Support vector machine**

The SVM method has wide applications in machine learning and pattern recognition. This classifier has the ability to deal with large, nonlinear data. SVM with kernel creates appropriate nonlinear decision boundaries to separate nonlinear data. In this method, the data is written to a high-dimensional space where the data is propagated in such a way that a linear decision boundary cloud separate it. The decision function for a kernel-based SVM is defined as follows

$$g(x) = sgn\left(Cov_{i=1}^T \alpha_i c_i k(x.x_i) + b\right) \qquad (14)$$

Where, $x = [1.\cdots.T]$ is a set of training instances, C represents the tags of the stack class $\alpha_i \geq 0$, a Lagrangian coefficient that is the result of a quadratic optimization problem. $k(x.x_i)$ represents the kernel and b is the bias. Selecting the kernel and setting the parameter values are important steps in designing an SVM classifier. The Gaussian RBF kernel has been selected for testing in this paper, which is defined as follows

$$k(x.x_i) = e^{-\gamma\|x-x_i\|^2} \qquad (15)$$

Where $\gamma = \frac{1}{2\sigma^2} > 0$ controls the width of the Gaussian function and $\|x - x_i\|$ is the middle term of x. In addition, SVM is overly in-sensitive to over-training, making it suitable for a variety of applications [25].

**2-3-3. k nearest neighbor**

K nearest neighbor is in the family of supervised learning algorithms. KNN is a sample-based learning algorithm that calculates the distance of the point we wish to label with the nearest points in classification according to the value specified for K. In addition, KNN decides on the label of the desired point according to the maximum number of these neighboring points. Various methods might be utilized to calculate this distance. Euclidean distance is one of the common ones

$$distance = \sqrt{\sum_i (x_i - y_i)^2} \qquad (16)$$

Accordingly, it chooses the k nearest samples from the training data, and then takes the majority vote of its classes. The most important parameter influencing the result of classification and labelling is determining the value of K in this algorithm, which changes the sample label as it changes [32].

How can the best value be considered for K to make a more accurate prediction? The answer to this question is that the determination of the value of K depends entirely on the data being examined, while the analyst determines the best value in term of accuracy by using different values of K.

## 2-4. Validation

A method for choosing between different models is needed when training a machine learning model. A common measure to evaluate a model performance is to measure the accuracy of unseen data. However, there could be a relatively large variation in the validation set of small amounts of data. There are different validation methods such as k-fold, random subsampling, etc. In the random subsampling method, some data are randomly selected to form test data. The rest of the data is also used for training. The model error rate in this method is also equal to the average error rate per repetition [33].

## 2-5 Proposed method

In this paper, Band-pass Butterworth filter pre-processing and CSP, TRCSP, FBCSP feature extraction methods along with LDA, KNN, SVM linear, and nonlinear classification have been studied. The steps of the approach are shown in Figure 3.

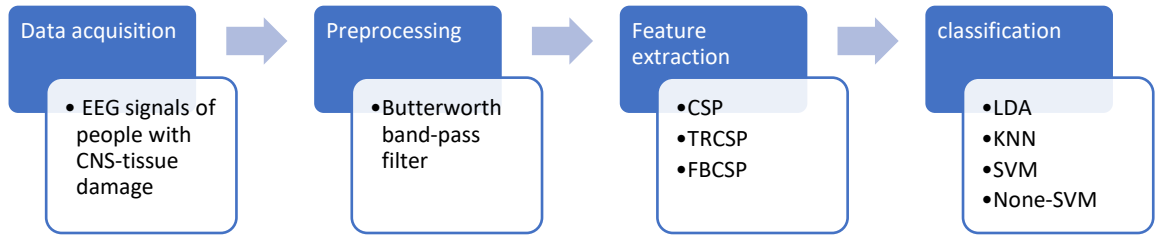

Figure3: Experimentation steps in this article

The accuracy measure is used to evaluate the EEG signal classification models in the BCI domain [3] as depicted in Equation

$$\text{Accuracy} = \frac{TP + TN}{TP + TN + FP + FN} \tag{17}$$

TP stands for true positive, TN for true negative, FP for false positive, and FN for false negative. Accordingly, the data related to 9 people with central nervous system (CNS) tissue damage are initially pre-processed by a Band-pass Butterworth filter with 8-30 Hz. Then the 5 mental tasks are reduced to desired two classes, and then the data are divided into train and test groups in a ratio of 70 to 30. Figure4 illustrates the data after these processes for entering the next step.

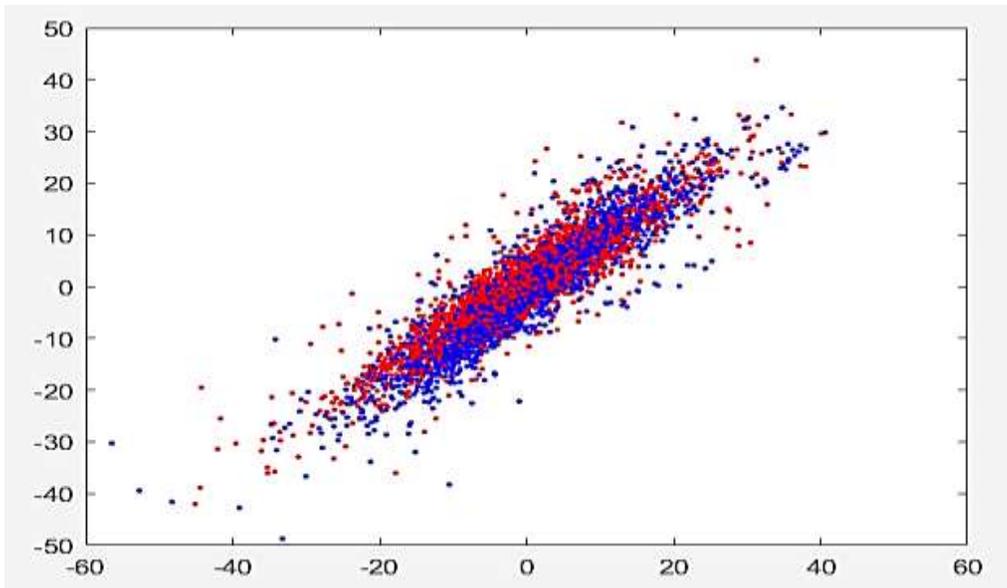

Figure4: EEG signals of patient A of classes WORD and FEET before feature extraction step. The red and blue points refer to classes WORD and FEET, respectively

CSP, TRCSP, and FBCSP methods are used for feature extraction. Figure 5 shows the feature extraction step.

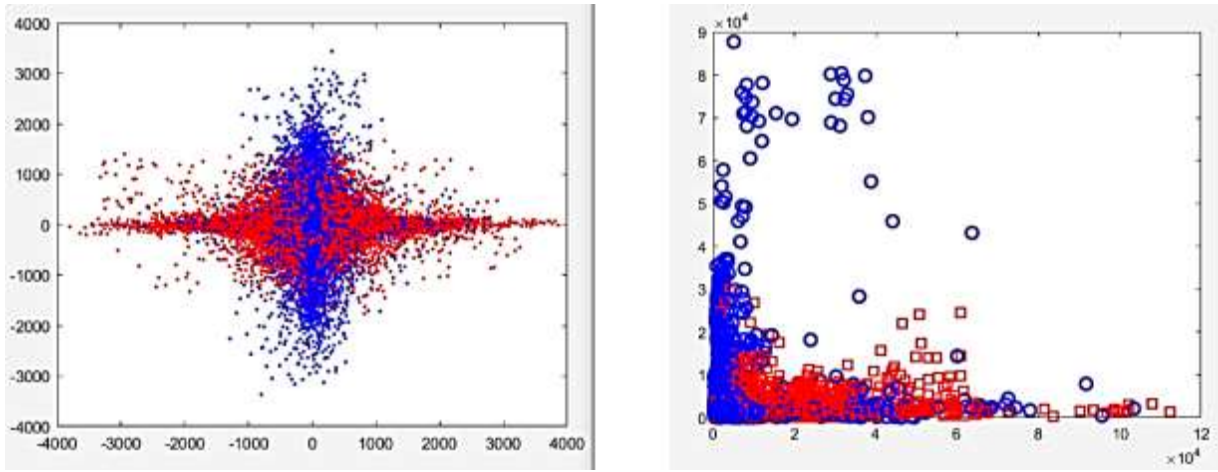

Figure5: EEG signals of patient A of classes WORD and FEET after FBCSP feature extraction step. The left figure shows the common spatial filtering on the data and the right figure is the variance of left one.

KNN, LDA and linear, and nonlinear SVM methods are used for classification. The training and test process is repeated up to 100 times and the random subsampling validation method is used to select the training and test data. The average accuracy per 100 repetitions is reported as the accuracy of the final result.

# 3. Results

In the experiments, data of five mental tasks received out of nine people with central nervous system (CNS) tissue damage were initially preprocessed with a Band-pass Butterworth filter in the range of 8-30 Hz to eliminate noise. The CSP, TRCSP, FBCSP algorithm for features extraction and KNN with k =1 and LDA and linear SVM and nonlinear SVM with RBF kernel for classification have been used for all possible binary classification of 5 mental tasks. The best results for each classifier for CSP, TRCSP, FBCSP algorithm are summarized in Table 2, table 3 and table 4 respectively. As it is shown, the rows indicate the classifier name while the second column is related to the binary combination of mental tasks with the highest accuracy among the other activities. Also, the third column contains the amount of classification accuracy for the subjects. Forth column shows the mean accuracy among the nine individuals with the standard deviation.

Table 2- Best Classification Results by CSP Method and Linear and Nonlinear Linear SVM, LDA, and KNN Classifiers

| classifier | action | accuracy | | | | | | | | | MEAN±STD |
|---|---|---|---|---|---|---|---|---|---|---|---|
| | | H | L | G | J | F | E | D | C | A | |
| SVM | Word vs hand | 51 | 50 | 51 | 50 | 54 | 50 | 51 | 51 | 51 | 51±1.0 |
| Non-SVM | Word vs hand | 55 | 51 | 53 | 49 | 54 | 49 | 50 | 49 | 53 | 54±2.3 |
| LDA | Nav and feet | 48 | 49 | 54 | 55 | 50 | 57 | 50 | 50 | 52 | 52±2.9 |
| KNN | Word vs feet | 64 | 67 | 67 | 72 | 68 | 68 | 67 | 69 | 72 | 68±1.0 |

Table 3- Best Classification Results by TRCSP Method and Linear and Nonlinear Linear SVM, LDA, and KNN Classifiers

| classifier | action | accuracy | | | | | | | | | MEAN±STD |
|---|---|---|---|---|---|---|---|---|---|---|---|
| | | H | L | G | J | F | E | D | C | A | |
| SVM | Sub vs feet | 50 | 50 | 52 | 51 | 52 | 49 | 51 | 50 | 50 | 51±0.9 |
| Non-SVM | Word vs nav | 54 | 51 | 55 | 51 | 57 | 53 | 52 | 50 | 52 | 53±2.2 |
| LDA | Word vs nav | 52 | 49 | 48 | 52 | 60 | 55 | 52 | 50 | 49 | 52±3.7 |
| KNN | Word vs feet | 65 | 67 | 67 | 73 | 67 | 68 | 67 | 69 | 73 | 68±2.0 |

Table4- Best Classification Results by FBCSP Method and Linear and Nonlinear Linear SVM, LDA, and KNN Classifiers

| classifier | action | accuracy | | | | | | | | | MEAN+_STD |
|---|---|---|---|---|---|---|---|---|---|---|---|
| | | H | L | G | J | F | E | D | C | A | |
| SVM | Word vs sub | 51 | 49 | 49 | 50 | 53 | 50 | 51 | 50 | 51 | 51±1.1 |
| Non-SVM | Word vs hand | 55 | 50 | 51 | 53 | 52 | 54 | 54 | 50 | 51 | 53±1.6 |
| LDA | Word vs nav | 52 | 50 | 48 | 51 | 60 | 54 | 51 | 50 | 50 | 52±3.7 |
| KNN | Word vs feet | 65 | 68 | 68 | 73 | 68 | 67 | 67 | 70 | 71 | 69±2.3 |

The results of evaluating of 9 individuals for all binary modes of activities are shown in figures 6,7,8. The row shows the binary combination of mental tasks of 9 patients, the column indicates the mean accuracy.

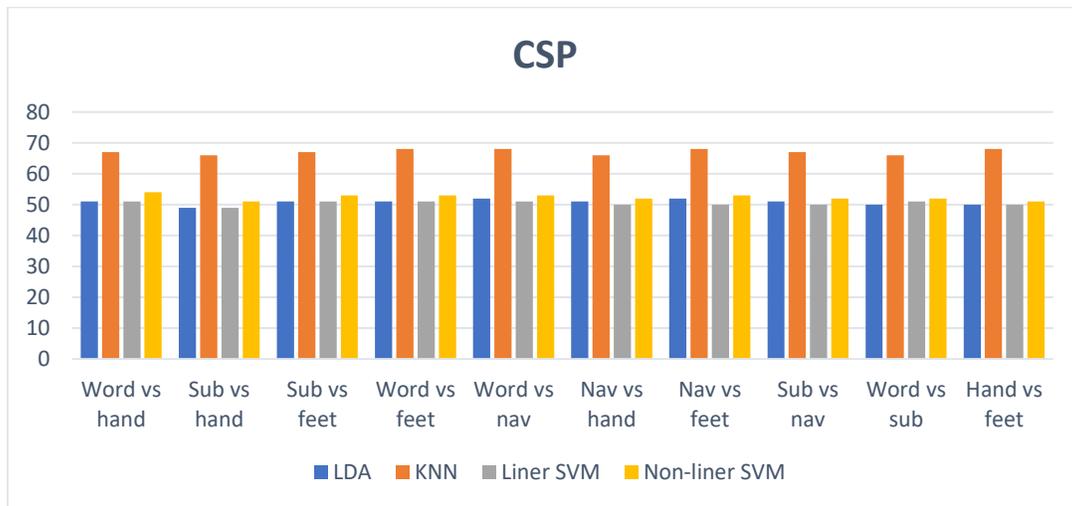

Figure6: CSP combination with LDA and KNN,

Linear and Nonlinear Linear SVM Classifiers for all binary modes of activity of 9 patients

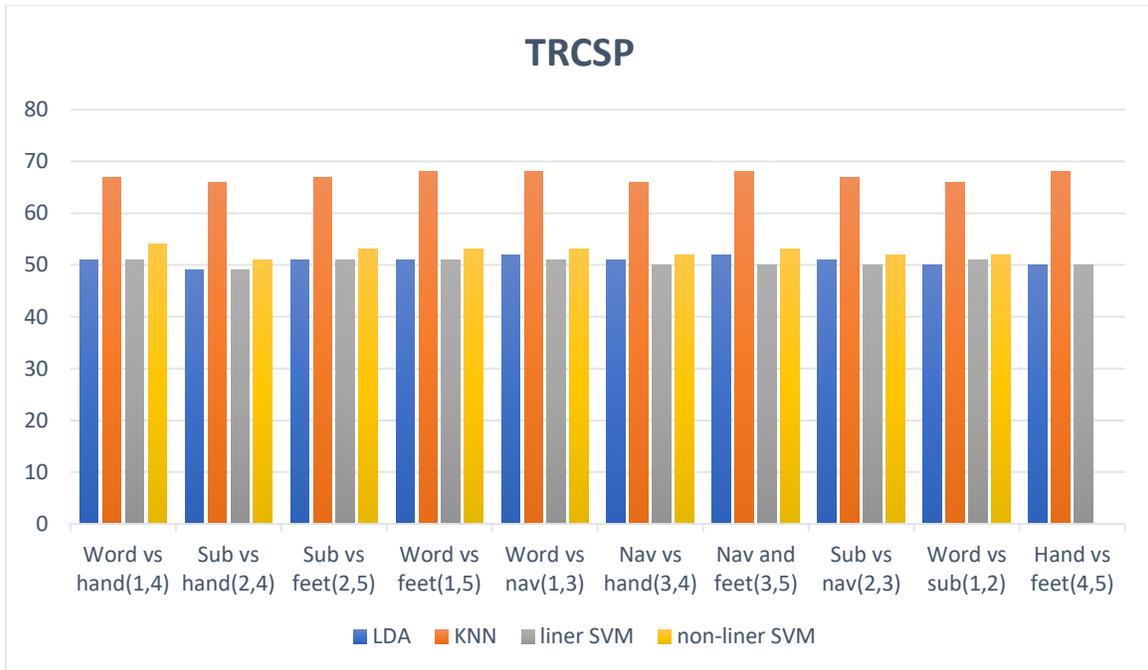

Figure7: TRCSP combination with LDA and KNN,

Linear and Nonlinear Linear SVM Classifiers for all binary modes of activity of 9 patients

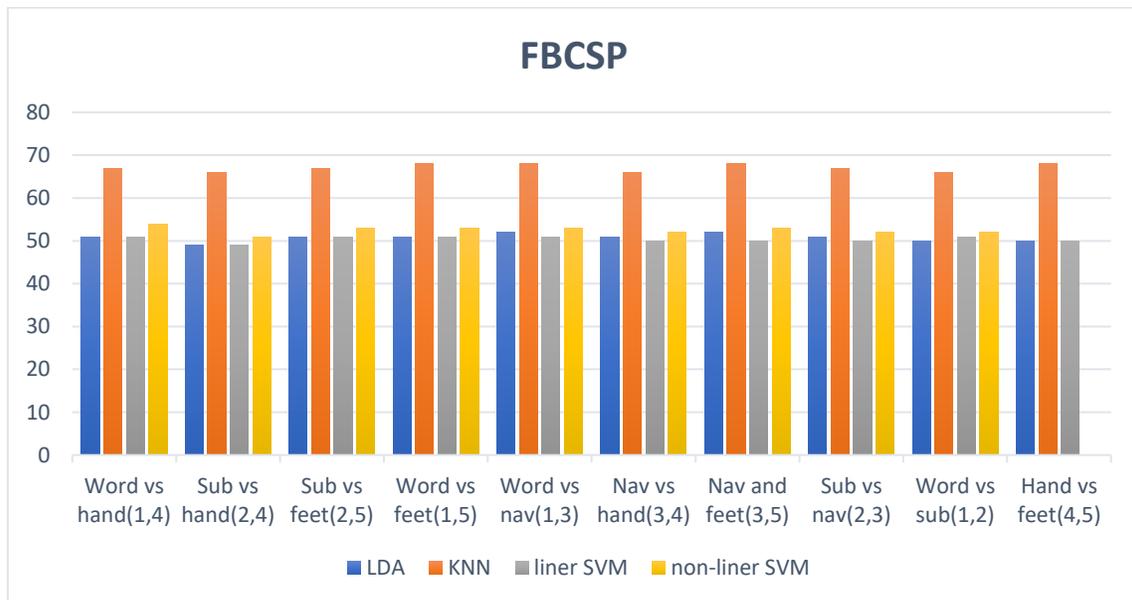

Figure8: FBCSP combination with LDA and KNN,

Linear and Nonlinear Linear SVM Classifiers for all binary modes of activity of 9 patients

# 4 - Conclusion

This article addresses an approach to increase the accuracy of brain-computer interface systems that can help the improvement of the living conditions of people with disabilities. CSP, TRCSP, FBCSP methods, along with LDA, KNN, linear and nonlinear SVM classifiers which have not been yet studied in the classification of this dataset, have been used to increase the accuracy of classification of EEG signals derived from mental tasks in users with the central nervous system (CNS) tissue damage. The accuracy of classification in all experiments changed significantly with the proper selection of the classifier. KNN achieved the highest accuracy among the other classifiers in all experiments as shown in figures 6, 7, and 8. However, no significant change in accuracy was obtained in the feature extraction section by changing the method. Only the FBCSP method had slightly better accuracy than the other two methods. Refer to figure 5, it could be found that the liner classification couldn't properly classify these data whilst the other classifiers like KNN based on distribution of data in figure5 should examined which results attested this.

The results showed that the classifiers' effectiveness in increasing the accuracy is better than the effectiveness of the feature extraction. The LDA and SVM classifiers are the two most popular classifiers, and the KNN classifier is less commonly used. This classifier has been unsuccessful on some databases compared to other classifiers. On the other hand, it has been more successful in some databases than different popular classifiers. Since the type of data studied in this article was similar to the databases on which the KNN classifier was successful, this classifier was used for classification and had excellent results compared to the other classifiers.

Overall, the 8-30Hz Butterworth band-pass preprocess by FBCSP feature extraction using the KNN classifier in two classes of word association and feet motor imagery achieved the highest accuracy in binary decoding among five mental tasks in people with the central nervous system (CNS) tissue damage. Also, motor imagery is commonly used for brain-computer interface systems. In practice, it has been observed that the combination of a mental task along with motor imagery has a higher accuracy for classification. This conclusion confirmed the discussion of mental tasks that can be used as well as motor imagery [22]; However, the accuracy of classification in the above article method was low. The methods used in this article increased the accuracy result while this increase was observed in a balanced way in all classes.

The patients studied in this article are classified in three groups of hemorrhagic strokes, locking syndrome, and spinal cord injury A and C. The classification results for the people in these groups were very close to each other. Since people with stroke suffer brain tissue damage and people with spinal cord injury have perfectly healthy brains, it can be concluded that the above method can decode the mental tasks of these three groups equally. As a result, this method can be generalized into 3 groups.

We face various challenges in EEG signal classification. The main challenges are the noise of signal, high-dimensional signal, costly and time-consuming data collection, significant differences in brain activity between individuals and even between sessions for one person. Data shortage, especially medical data, feature extraction, and classification accuracy are among the serious challenges in this area. Finding shows a need to increase the accuracy in classifying the mental tasks of these people to be used in virtual reality systems, the Internet of things, and various types of brain-computer interface such as communication, movement, movement control, and rehabilitation for practical implementation in the real world.

Clinical Significance

People with central nerve damage, such as those with stroke and spinal cord injury who cannot communicate with their surroundings, need to be cared for regularly. This is costly and time-consuming. The basic needs of people with spinal cord injuries who cannot normally communicate with their surroundings can be identified correctly and quickly by increasing the accuracy in brain activity classification and its use in brain-computer and communication interfaces. This meets their basic needs and thus improves their quality of life. Also, most of the studies in this field have been on healthy people, and few articles have examined the processing methods on the data of people with disabilities. Successful processing methods in this area have been done precisely on these people. Providing the results of processing on disabled people has opened future research to examine the effects of processing on healthy people. Since the patients studied in this article suffered from spinal cord injury, hemorrhagic stroke, locking syndrome, the results, and methods used in the article in brain computer and communication systems can be helpful for these people. The output of this work can be used in IoT systems designed to connect these people with the environment. Consequently, the idea of using successful methods in signal processing to increase the accuracy and solve the problem of data shortage on such systems seems necessary. It is also recommended to apply various pre-processors such as wavelets, etc. on the data to investigate their effect on future work accuracy. Also, other successful methods for classifying such as neural network and its expansion like deep learning are proposed. Multi-class classification is suggested to extend the capabilities of brain-computer interface systems. Since people with disabilities are hopeful and interested in life even with the most severe type of disability, we should try to provide better conditions for them and pay more attention to studies that improve their living conditions.